\documentclass{article}

    \usepackage[preprint, nonatbib]{neurips_2020}

\usepackage[utf8]{inputenc} 
\usepackage[T1]{fontenc}    
\usepackage{hyperref}       
\usepackage{url}            
\usepackage{booktabs}       
\usepackage{amsfonts}       
\usepackage{nicefrac}       
\usepackage{microtype}      
\usepackage{amssymb}
\usepackage{amsmath}
\usepackage{bm}
\usepackage{enumitem}
\usepackage{amssymb}
\setlist{nosep}

\newcommand{\etc}{etc}

\title{Detecting Individuals with Depressive Disorder from Personal Google Search and YouTube History Logs}

\author{%
   Boyu Zhang\\
   Department of Computer Science\\
   University of Rochester\\
   Rochester, NY \\
   \texttt{bzhang25@u.rochester.edu} \\
   \And
   Anis Zaman\\
   Department of Computer Science\\
   University of Rochester\\
   Rochester, NY \\
   \texttt{azaman2@cs.rochester.edu} \\
   \And
   Rupam Acharyya\\
   Department of Computer Science\\
   University of Rochester\\
   Rochester, NY \\
   \texttt{racharyy@cs.rochester.edu} \\
   \And
   Ehsan Hoque \\
   Department of Computer Science\\
   University of Rochester\\
   Rochester, NY \\
   \texttt{mehoque@gmail.com} \\
   \And 
   Vincent Silenzio\\
   Department of Urban-Global Public Health\\
   Rutgers University\\
   Jersey City, NJ \\
   \texttt{vincent.silenzio@rutgers.edu} \\
   \And
   Henry Kautz\\
   Department of Computer Science\\
   University of Rochester\\
   Rochester, NY \\
   \texttt{henry.kautz@gmail.com} \\
}

\begin{document}

\maketitle

\begin{abstract}
Depressive disorder is one of the most prevalent mental illnesses among the global population. However, traditional screening methods require exacting in-person interviews and may fail to provide immediate interventions. In this work, we leverage ubiquitous personal longitudinal Google Search and YouTube engagement logs to detect individuals with depressive disorder. We collected Google Search and YouTube history data and clinical depression evaluation results from $212$ participants ($99$ of them suffered from moderate to severe depressions). We then propose a personalized framework for classifying individuals with and without depression symptoms based on mutual-exciting point process that captures both the temporal and semantic aspects of online activities. Our best model achieved an average F1 score of $0.77 \pm 0.04$ and an AUC ROC of $0.81 \pm 0.02$.  
\end{abstract}

\section{Introduction and Related Work}
According to the National Institute of Health, it is estimated that more than $17$ million adults in the United States have at least one major depressive episode every year. This number represents $7.1\%$ of all U.S. adults\footnote{\url{https://www.nimh.nih.gov/health/statistics/major-depression.shtml}}. Nonetheless, very few patients in need received immediate and proper medical interventions \cite{kessler2005prevalence, wang2005failure}. The prevalence of mental illness has become one of the most significant burden for the economy and human well-beings in the U.S. \cite{whiteford2013global}.

Traditional care delivery methods have failed to ameliorate the rampant depression problems among large populations. The impeded help delivery is mostly due to the exacting traditional screening approaches such as in-person interviews. The current healthcare system requires patients to actively reach out to caregivers and be physically presented in clinics for assessments. However, such practice may be blocked by time availability, expenses, and the unawareness of the patients. Moreover, the diagnosis is prone to concealing information and social stigmas as the patients may not be willing to reveal all personal details, especially among teenagers \cite{eisenberg2009stigma, hunt2010mental}.

A non-invasive technique in mental health surveillance and intervention can be based on online ubiquitous data. As many individuals spend their lives online every day for a considerable amount of time, the digital footprints left behind may capture the cognitive and mental states of mind of the user at different moments. Most importantly, these digital traces may preserve information that is useful in flagging users at risks of mental health problems. Extensive researches have probed pervasive online data for various mental illnesses. Reddit \cite{shen2017detecting, fraga2018online, gaur2018let} and Twitter \cite{de2013predicting} have been explored to detect anxiety and depression. \cite{seabrook2016social} provided a comprehensive review on utilizing social network sites to examine anxiety and depressive disorders. More detailed evaluations of depression targeting young adults were also performed, as in \cite{lin2016association}, and a positive correlation was found between social media usages and depressive disorders.

Yet, as it has long been critically addressed, public online platforms are subject to self-censorship where users with mental health difficulties may refrain from generating contents due to peer stigmas, creating a false negative image. Besides, many of the above work merely detects population-level mental disorders but fails to establish a personalized healthcare model that is more clinically meaningful. To tackle these problems, some studies have investigated building individual-level mental health tracking systems from private online data. One promising data source is personal search history, and it has been applied in detecting low self-esteem \cite{zaman2019detecting} and schizophrenia spectrum disorders \cite{birnbaum2020utilizing} among young individuals. \cite{zaman2020estimating} furthered the experiments with YouTube histories to both detect and predict anxiety disorders among college students. 

Inspired by these results, this study focuses on the task of depressive disorder detection with a similar data collection pipeline as \cite{zaman2020estimating}. However, instead of relying on explicit predefined labels for search logs and YouTube videos through the Google NLP API\footnote{\url{https://cloud.google.com/natural-language/docs/classify-text-tutorial}} or LIWC \cite{tausczik2010psychological}, we leverage distributional phrase embeddings for semantic information. Furthermore, different from \cite{zaman2019detecting, zaman2020estimating, birnbaum2020utilizing} where the researchers utilized rule-based temporal features such as hourly activity counts and late night engagements, we exploit the potential of multidimensional mutual-exciting point processes in integrating the stochastic and semantic aspects of online activities. The intuition is that online activities with distinct semantics may trigger one another at different times, and this may characterize the behaviors of users with and without depression symptoms. Our best model achieved an average weighted F1 score of $0.77 \pm 0.04$ and an AUC ROC of $0.81 \pm 0.02$ in the depression classification task. 

\section{Data}
Our data consists of two parts: i) personal longitudinal Google Search and YouTube engagement logs and ii) Patient Health Questionnaire-9 depression survey responses (PHQ-9) \cite{kroenke2001phq} from the participants. Similar to \cite{zaman2020estimating}, we utilized the Google Takeout\footnote{\url{http://takeout.google.com/}} platform to collect individual-level Google Search and YouTube engagement data. Each participant must be 18-year-old with an active Google account to qualify for the study. The PHQ-9 survey responses were collected via face-to-face interviews, and a final score was calculated from each participant. Given the proprietary nature of the data collected and the safety of human participants involved, our data collection pipeline has been thoroughly vetted and approved by our Institutional Review Board (IRB).

In total, $223$ participants volunteered in the study, and $212$ of them provided valid online data and PHQ-9 responses. $98\%$ of the participants were undergraduate students, and the rest $2\%$ were graduate student. All of them came from the same college in the U.S. $69\%$ of the participants were female, and $28\%$ of them were male. The rest $3\%$ reported non-binary genders.  

\subsection{Depressive Disorder Measurements}\label{depression_measurement_section}
The PHQ-9 score can range from $0$ to $27$, and a score $\geq 15$ is considered as moderate to severe depressions that may require medical interventions \cite{kroenke2001phq, kroenke2012enhancing}. Thus, we set the cutoff value at $15$. We labeled the participants with a PHQ-9 score $\geq 15$ as the \textit{Depressed} group and those with a score $< 15$ as the \textit{Healthy} group. Out of the $212$ volunteer participants, $99$ ($46.7\%$) of them belongs to the \textit{Depressed} group, and the rest $113$ ($53.3\%$) of them belongs to the \textit{Healthy} group. 

\subsection{Online History Data}
For Google Search and YouTube data, the participants zipped their longitudinal online logs via the Google Takeout platform and shared with the research team. All contents that could lead to malicious linkage attacks \cite{dwork2014algorithmic} or reveal personal identities, such as name, contacts, GPS locations, and financial information, were obscured and removed by the Data Loss Prevention (DLP) API~\cite{kiang2016data,kim2016cloud} from Google before the research team analyzed data. It is worth mentioning that the Google Takeout platform records and archives all the search and YouTube histories associated with each Google account. Therefore, no matter what device the participant was using (smartphones, iPad, laptops, \etc), the online history would be included in the data as long as the participant was logged in with his/her/their accounts.

Each activity in Google Search and YouTube engagement logs has a timestamp to the precision of seconds. For Google Search, the engagement log contained the query input by the user. For YouTube, the engagement log contained the URL to the videos watched by the participants. We further retrieved the meta-data of the videos watched through the official YouTube API\footnote{\url{https://developers.google.com/youtube/v3/docs}}, including the title, duration, and the numbers of likes and dislikes. The Google Search logs and YouTube logs were merged together chronologically.

In total, we collected $1,989,372$ Google Search queries and $1,078,302$ YouTube watched videos from all participants. On average, the online history log spans around $5.1$ years for every person. 

\section{Model: classifying individuals with depressive disorder}
First, we obtain a semantic embedding for each Google Search and YouTube video in online history logs. We pass each search query and video title through BERT base model \cite{devlin2018bert} and retrieve the CLS token ($\in \mathbb{R}^{768}$) as a vector representation. This procedure is done for the search queries and YouTube videos from all participants. 

Next, we cluster all the online activity embeddings and identify $K$ centers as implicit \textit{topics} through k-means. Thereby, each Google Search and YouTube video can be labeled with a topic based on the cluster it belongs to. After that, the longitudinal online history log of each user can be described as a series of timestamped events with topic labels. This is further formally denoted as a marked temporal point process \cite{aalen2008survival} with $K$ mark choices. Such a data-driven topic modeling process avoids predefined explicit labels such as the Google NLP categories used in \cite{zaman2019detecting}, which may limit the feature space and lose a considerable amount of signals. 

Then, for each participant, we fit a $K$-dimensional Hawkes process with an exponential decay kernel, and each dimension corresponds to a topic. The intensity $\lambda_i(t)$ of the Hawkes process is defined as:
\begin{equation}
    \forall i \in [1, ..., K], \lambda_i(t) = \mu_i + \sum_{j = 1}^{K}\sum_{m : t_{m}^{j} < t}\alpha^{ij}\beta^{ij}\exp\left(-\beta^{ij}(t - t_{m}^{j})\right) 
\end{equation}
where $\mu_i$ is the baseline intensity for topic $i$, $\alpha^{ij}$ is the expected number of events in topic $i$ excited by a previous event in topic $j$, and $\beta^{ij}$ is the decay of intensity for topic $i$ following an event in $j$. This mutual-exciting point process captures the stochastic nature of online activities from different topics. Due to the well known non-convex problem \cite{lewis2011nonparametric, zhou2013learning}, we only optimize the baseline intensity $\boldsymbol{\mu} \in \mathbb{R}^K$ and the adjacency matrix $\boldsymbol{\alpha} \in \mathbb{R}^{K \times K}$. Decay rates $\boldsymbol{\beta} \in \mathbb{R}^{K \times K}$ are predefined such that the intensity decays slower when topic $i$ and $j$ are similar. Specifically, we take all the topic cluster centers $\varphi_i$ and employ a RBF kernel $\exp\left(-\|\varphi_i - \varphi_j\|^2 / \sigma^2 \right)$ to measure the similarity. The BERT embeddings provide semantic information and implicit topics for Google Search and YouTube videos, and the multidimensional Hawkes process captures the temporal interplay between online engagements. We envision that the stochastic nature of different online activities characterizes the user behavior and may be useful in distinguishing between the \textit{Depressed} and \textit{Healthy} groups. 

After the optimization, we obtain a pair of [$\boldsymbol{\alpha}$, $\boldsymbol{\mu}$] for each user. While $\boldsymbol{\mu}$ can be used as a personalized feature vector directly, $\boldsymbol{\alpha}$ is a weighted adjacency matrix. By viewing the $\boldsymbol{\alpha}$ as a directed weighted graph, we are interested in, for each topic (vertex) $i$, the incoming weights versus the outgoing weights. Thus, we calculate $\boldsymbol{\phi} \in \mathbb{R}^K$ such that $\forall i \in [1, ..., K], \phi_{i} = \sum_{j}\alpha^{ij} / \sum_{j}\alpha^{ji}$ and use it as another feature vector. 

On the one hand, it has been reported that depressive disorder shall be diagnosed with symptoms lasting for at least two
weeks\footnote{\url{https://www.nimh.nih.gov/health/topics/depression/index.shtml}}. On the other hand, while we have a fairly rich online history log for each person ($5.1$ years on average), we suspect few participants would have persistent depression symptoms for five years. An online history log that is too long or too short may both lose crucial behavioral signals about the depressive disorder of the user. Thus, when fitting a mutual-exciting point process for each person, we experimented with assorted durations $D$ of time spans and truncated the longitudinal online engagement logs accordingly in each round. Concretely, for any $D$ picked, we fitted the point process with the data $D$ months/weeks before we received the PHQ-9 survey responses, see Section \ref{results_section} for details. 

Finally, we feed $\boldsymbol{\phi}$ and $\boldsymbol{\mu}$, separately, as the input to a L2-regularized Support Vector Machine with a linear kernel to classify participants with and without moderate to severe depressive disorders. The binary labels and cutoff value are stated in Section \ref{depression_measurement_section}.

\section{Experiments and Results}\label{results_section}
\begin{table}
  \caption{The performance of $\boldsymbol{\mu}$ features.}
  \label{mu-table}
  \centering
  \begin{tabular}{lcccc}
    \toprule
     & Depressed & Healthy & Weighted Avg.\\
    \midrule
    Precision & $0.71 \pm 0.08$ & $0.69 \pm 0.04$ & $0.70 \pm 0.05$\\
    Recall    & $0.80 \pm 0.06$ & $0.57 \pm 0.17$ & $0.70 \pm 0.05$\\
    F1 score  & $0.75 \pm 0.04$ & $0.61 \pm 0.10$ & $0.69 \pm 0.06$\\
    \midrule
    AUC ROC   & \multicolumn{3}{c}{$0.74 \pm 0.04$} \\
    \bottomrule
  \end{tabular}
\end{table}
\begin{table}
  \caption{The performance of $\boldsymbol{\phi}$ features.}
  \label{phi-table}
  \centering
  \begin{tabular}{lcccc}
    \toprule
     & Depressed & Healthy & Weighted Avg.\\
    \midrule
    Precision & $0.88 \pm 0.02$ & $0.71 \pm 0.04$ & $0.81 \pm 0.04$\\
    Recall    & $0.82 \pm 0.03$ & $0.79 \pm 0.06$ & $0.82 \pm 0.05$\\
    F1 score  & $0.80 \pm 0.03$ & $0.73 \pm 0.03$ & $\boldsymbol{0.77 \pm 0.04}$\\
    \midrule
    AUC ROC   & \multicolumn{3}{c}{$\boldsymbol{0.81 \pm 0.02}$} \\
    \bottomrule
  \end{tabular}
\end{table}

We carried out the experiments in several 5-fold cross-validations with various hyperparameters. The hyperparameters include: i) the number of implicit topics $K$, ii) the duration $D$ of online data truncated when fitting the $K$-dimensional Hawkes process, iii) the $\sigma$ in the RBF kernel for the predefined decay rates, and iv) the regularization parameter $C$ in the SVM.

We grid searched through $K\in \{5, 10, 15, 20, 25\}$, $\sigma\in \{0.001, 0.01, 0.1, 1, 10\}$, and $C\in \{0.1, 1, 10, 100\}$. We experimented with $D = 2$ weeks, $4$ weeks, $3$ months, $6$ months, $12$ months, and the whole data series. For each combination of the four hyperparameters, we performed a 5-fold cross-validation. In general, we found that, when $\sigma = 0.01$ and $C = 1$, the performance was relatively the best for each pair of $K$ and $D$. 

Across all the groups, the best performance was achieved with $\sigma = 0.01$, $C = 1$, $K = 10$, and $D = 6$ months, and we reported the detailed per-class and weighted averages in Table \ref{mu-table} and \ref{phi-table} for the baseline intensity features $\boldsymbol{\mu}$ and processed vertex weight features $\boldsymbol{\phi}$ , respectively. The $\boldsymbol{\phi}$ features achieved the best average weighted F1 score of $0.77 \pm 0.04$ and AUC ROC of $0.81 \pm 0.02$ in discriminating between the two groups. 


\section{Discussion}
In this work, we have shown that personal Google Search and YouTube histories can provide robust behavioral representations for classifying users with and without depressive disorder. By utilizing multidimensional Hawkes processes and distributional semantic embeddings, we are able to capture the interplay between activities of different implicit topics. 

Yet, there are many limitations. First, given the sensitive nature of the longitudinal data collected, there remains significant obstacles for real-world applications such as data privacy and the safety of the participants. Throughout our study, the volunteer participants reserve the rights to opt-out and remove their data at any time. Also, our data storage is cloud-based and HIPAA-compliant. Moreover, a clinical decision made fully by automated computation systems rises ethical concerns inevitably. We envision such a system to take up an assisting role, at most, in offering medical suggestions. The ultimate judgement should always be made by experts who fully understand both the medical knowledge and the limitations of the models. At last, this study only focused on college students, and further investigations are required to assess the robustness of the model across populations and backgrounds.

\bibliography{ref}

\begin{thebibliography}{10}

\bibitem{aalen2008survival}
Odd Aalen, Ornulf Borgan, and Hakon Gjessing.
\newblock {\em Survival and event history analysis: a process point of view}.
\newblock Springer Science \& Business Media, 2008.

\bibitem{birnbaum2020utilizing}
Michael~Leo Birnbaum, Anna Van~Meter, Victor Chen, Asra~F Rizvi, Elizabeth
  Arenare, Munmun De~Choudhury, John~M Kane, et~al.
\newblock Utilizing machine learning on internet search activity to support the
  diagnostic process and relapse detection in young individuals with early
  psychosis: Feasibility study.
\newblock {\em JMIR Mental Health}, 7(9):e19348, 2020.

\bibitem{de2013predicting}
Munmun De~Choudhury, Michael Gamon, Scott Counts, and Eric Horvitz.
\newblock Predicting depression via social media.
\newblock {\em Icwsm}, 13:1--10, 2013.

\bibitem{devlin2018bert}
Jacob Devlin, Ming-Wei Chang, Kenton Lee, and Kristina Toutanova.
\newblock Bert: Pre-training of deep bidirectional transformers for language
  understanding.
\newblock {\em arXiv preprint arXiv:1810.04805}, 2018.

\bibitem{dwork2014algorithmic}
Cynthia Dwork, Aaron Roth, et~al.
\newblock The algorithmic foundations of differential privacy.
\newblock {\em Foundations and Trends in Theoretical Computer Science}, 2014.

\bibitem{eisenberg2009stigma}
Daniel Eisenberg, Marilyn~F Downs, Ezra Golberstein, and Kara Zivin.
\newblock Stigma and help seeking for mental health among college students.
\newblock {\em Medical Care Research and Review}, 66(5):522--541, 2009.

\bibitem{fraga2018online}
Barbara~Silveira Fraga, Ana Paula~Couto da~Silva, and Fabricio Murai.
\newblock Online social networks in health care: A study of mental disorders on
  reddit.
\newblock In {\em 2018 IEEE/WIC/ACM International Conference on Web
  Intelligence (WI)}, pages 568--573. IEEE, 2018.

\bibitem{gaur2018let}
Manas Gaur, Ugur Kursuncu, Amanuel Alambo, Amit Sheth, Raminta Daniulaityte,
  Krishnaprasad Thirunarayan, and Jyotishman Pathak.
\newblock " let me tell you about your mental health!" contextualized
  classification of reddit posts to dsm-5 for web-based intervention.
\newblock In {\em Proceedings of the 27th ACM International Conference on
  Information and Knowledge Management}, pages 753--762, 2018.

\bibitem{hunt2010mental}
Justin Hunt and Daniel Eisenberg.
\newblock Mental health problems and help-seeking behavior among college
  students.
\newblock {\em Journal of adolescent health}, 46(1):3--10, 2010.

\bibitem{kessler2005prevalence}
Ronald~C Kessler, Wai~Tat Chiu, Olga Demler, and Ellen~E Walters.
\newblock Prevalence, severity, and comorbidity of 12-month dsm-iv disorders in
  the national comorbidity survey replication.
\newblock {\em Archives of general psychiatry}, 62(6):617--627, 2005.

\bibitem{kiang2016data}
Andy Kiang and Joel Bailon.
\newblock Data loss prevention (dlp) methods and architectures by a cloud
  service, January~12 2016.
\newblock US Patent 9,237,170.

\bibitem{kim2016cloud}
Tae~Wan Kim and Seung~Tae Paek.
\newblock Cloud data discovery method and system for private information
  protection and data loss prevention in enterprise cloud service environment,
  October~27 2016.
\newblock US Patent App. 14/728,503.

\bibitem{kroenke2012enhancing}
Kurt Kroenke.
\newblock Enhancing the clinical utility of depression screening.
\newblock {\em CMAJ}, 184(3):281--282, 2012.

\bibitem{kroenke2001phq}
Kurt Kroenke, Robert~L Spitzer, and Janet~BW Williams.
\newblock The phq-9: validity of a brief depression severity measure.
\newblock {\em Journal of general internal medicine}, 16(9):606--613, 2001.

\bibitem{lewis2011nonparametric}
Erik Lewis and George Mohler.
\newblock A nonparametric em algorithm for multiscale hawkes processes.
\newblock {\em Journal of Nonparametric Statistics}, 1(1):1--20, 2011.

\bibitem{lin2016association}
Liu~Yi Lin, Jaime~E Sidani, Ariel Shensa, Ana Radovic, Elizabeth Miller,
  Jason~B Colditz, Beth~L Hoffman, Leila~M Giles, and Brian~A Primack.
\newblock Association between social media use and depression among us young
  adults.
\newblock {\em Depression and anxiety}, 33(4):323--331, 2016.

\bibitem{seabrook2016social}
Elizabeth~M Seabrook, Margaret~L Kern, and Nikki~S Rickard.
\newblock Social networking sites, depression, and anxiety: a systematic
  review.
\newblock {\em JMIR mental health}, 3(4):e50, 2016.

\bibitem{shen2017detecting}
Judy~Hanwen Shen and Frank Rudzicz.
\newblock Detecting anxiety through reddit.
\newblock In {\em Proceedings of the Fourth Workshop on Computational
  Linguistics and Clinical Psychology—From Linguistic Signal to Clinical
  Reality}, pages 58--65, 2017.

\bibitem{tausczik2010psychological}
Yla~R Tausczik and James~W Pennebaker.
\newblock The psychological meaning of words: Liwc and computerized text
  analysis methods.
\newblock {\em Journal of language and social psychology}, 29(1):24--54, 2010.

\bibitem{wang2005failure}
Philip~S Wang, Patricia Berglund, Mark Olfson, Harold~A Pincus, Kenneth~B
  Wells, and Ronald~C Kessler.
\newblock Failure and delay in initial treatment contact after first onset of
  mental disorders in the national comorbidity survey replication.
\newblock {\em Archives of general psychiatry}, 62(6):603--613, 2005.

\bibitem{whiteford2013global}
Harvey~A Whiteford, Louisa Degenhardt, J{\"u}rgen Rehm, Amanda~J Baxter,
  Alize~J Ferrari, Holly~E Erskine, Fiona~J Charlson, Rosana~E Norman,
  Abraham~D Flaxman, Nicole Johns, et~al.
\newblock Global burden of disease attributable to mental and substance use
  disorders: findings from the global burden of disease study 2010.
\newblock {\em The lancet}, 382(9904):1575--1586, 2013.

\bibitem{zaman2019detecting}
Anis Zaman, Rupam Acharyya, Henry Kautz, and Vincent Silenzio.
\newblock Detecting low self-esteem in youths from web search data.
\newblock In {\em The World Wide Web Conference}, pages 2270--2280, 2019.

\bibitem{zaman2020estimating}
Anis Zaman, Boyu Zhang, Vincent Silenzio, Ehsan Hoque, and Henry Kautz.
\newblock Estimating anxiety based on individual level engagements on youtube
  \& google search engine.
\newblock {\em arXiv preprint arXiv:2007.00613}, 2020.

\bibitem{zhou2013learning}
Ke~Zhou, Hongyuan Zha, and Le~Song.
\newblock Learning triggering kernels for multi-dimensional hawkes processes.
\newblock In {\em International Conference on Machine Learning}, pages
  1301--1309, 2013.

\end{thebibliography}







\end{document}